\documentclass[12pt]{article}          
\usepackage[margin=1in]{geometry}      
\usepackage{amsmath,amssymb,amsfonts}  
\usepackage{amsthm}                    
\usepackage{graphicx}                  
\usepackage{booktabs}                  
\usepackage{hyperref}                  
\usepackage{cleveref}                  
\usepackage{float}                     
\usepackage[round]{natbib}             
\title{Chaos Gated Tunneling Drives Molecular Reactivity in Astrophysical Environments}
\author{
  Saptarshi~G.~Dastider%
    \thanks{Department of Chemistry, Indian Institute of Space Science and Technology (IIST),\\
            Thiruvananthapuram 695547, Kerala, India; Email: saptarshhigdask@gmail.com}%
    \thanks{Corresponding author}
  \\[2pt]
  K.~Prashant%
    \thanks{Department of Chemistry, Indian Institute of Space Science and Technology (IIST),\\
            Thiruvananthapuram 695547, Kerala, India}%
  \\[2pt]
  P.~Shruti%
    \thanks{Department of Chemistry, Indian Institute of Space Science and Technology (IIST),\\
            Thiruvananthapuram 695547, Kerala, India}%
  \\[2pt]
  C.~Sudheesh%
    \thanks{Department of Chemistry, Indian Institute of Space Science and Technology (IIST),\\
            Thiruvananthapuram 695547, Kerala, India; Email: sudheesh@iist.ac.in}%
    \thanks{Corresponding author}
  \\[2pt]
  Jobin~Cyriac%
    \thanks{Department of Chemistry, Indian Institute of Space Science and Technology (IIST),\\
            Thiruvananthapuram 695547, Kerala, India; Email: jobincyriac@iist.ac.in}%
    \thanks{Corresponding author}
}\author{
  Saptarshi~G.~Dastider%
    \thanks{Department of Chemistry, Indian Institute of Space Science and Technology (IIST),\\
            Thiruvananthapuram 695547, Kerala, India; Email: saptarshhigdask@gmail.com}%
    \thanks{Corresponding author}
  \\[2pt]
  K.~Prashant%
    \thanks{Department of Physics, Indian Institute of Space Science and Technology (IIST),\\
            Thiruvananthapuram 695547, Kerala, India}%
  \\[2pt]
  P.~Shruti%
    \thanks{Department of Physics, Indian Institute of Space Science and Technology (IIST),\\
            Thiruvananthapuram 695547, Kerala, India}%
  \\[2pt]
  C.~Sudheesh%
    \thanks{Department of Physics, Indian Institute of Space Science and Technology (IIST),\\
            Thiruvananthapuram 695547, Kerala, India; Email: sudheesh@iist.ac.in}%
    \thanks{Corresponding author}
  \\[2pt]
  Jobin~Cyriac%
    \thanks{Department of Chemistry, Indian Institute of Space Science and Technology (IIST),\\
            Thiruvananthapuram 695547, Kerala, India; Email: jobincyriac@iist.ac.in}%
    \thanks{Corresponding author}
}
\date{}  

\begin{document}
\maketitle

\begin{abstract}
Accurate modeling of ion-molecule reaction networks is essential for understanding the chemical evolution of planetary ionospheres, particularly for giant planets where proton-transfer chains drive atmospheric composition. However, predicting reaction rates in these ultracold environments remains a challenge due to the non-trivial interplay between vibrational dynamics and quantum tunneling. In this work we present a chaos‑diagnostic framework that integrates multireference electronic structure theory, Adiabatic Gauge Potentials (AGP), and Random Matrix Theory (RMT) to characterize the microscopic dynamics of proton transport. Using the formation of H$_3^+$ and the proton‑bound cluster H$_5^+$ as representative model systems relevant to Jovian atmospheres, we demonstrate that the transition state acts as a dynamical bottleneck where quantum chaos is notably suppressed ($\langle r\rangle\approx0.36$), effectively enhancing tunneling probabilities. We introduce a “fragility index” based on the AGP slope to quantify how specific vibrational modes reintroduce chaos and suppress reactivity. This diagnostic approach offers a generalizable, data‑driven metric for identifying vibrationally gated pathways in complex astrochemical networks, providing a theoretical basis for refining kinetic models of planetary and interstellar plasmas.
\end{abstract}


The accurate modeling of planetary ionospheres, particularly those of gas giants like Jupiter, relies fundamentally on understanding the complex ion-molecule reaction networks driven by proton transfer. While H$_3^+$ is well-established as the universal proton donor and a primary tracer of cosmic-ray ionization in these environments\cite{Oka2006,Indriolo2007,Moore2019}, its subsequent interactions to form proton-bound clusters like H$_5^+$ play a critical role in collisional cooling and atmospheric composition\cite{Tennyson2019,Herbst2021}. Despite the centrality of these species to astrochemical networks\cite{Drossart2019,Valek2020}, the kinetic rates governing their formation in ultracold environments remain difficult to predict. This uncertainty stems largely from the non-trivial interplay between vibrational dynamics and quantum tunneling—a dominant mechanism in low-temperature interstellar clouds and planetary auroras\cite{Wild2023,Schreiner2020}.

Standard kinetic models often rely on Classical Transition State Theory (TST)\cite{IRC,Bell1980}, which emphasizes barrier heights and zero-point energy. However, TST cannot predict when the quantum coherence essential for tunneling is preserved or destroyed by vibrational excitations. Recent experimental evidence demonstrates mode-selective control of tunneling in ion-molecule reactions\cite{Wild2023}, while theoretical advances indicate that chaotic vibrational mixing can ``scramble'' quantum information, effectively suppressing tunneling even when energetically allowed\cite{Zhang2024_PNAS,Zhang2024_scrambling}. This suggests that the kinetic stability of atmospheric ions is governed not just by energetics, but by the underlying quantum chaotic dynamics.

To diagnose these dynamical effects, Quantum Chaos theory provides a powerful, yet underutilized, framework. While Random Matrix Theory (RMT) links chaotic classical systems to universal spectral statistics (Wigner-Dyson distribution)\cite{Wigner1955,Haake2010,Bohigas1984}, it lacks a direct connection to molecular geometry. The Adiabatic Gauge Potential (AGP)—which quantifies the sensitivity of eigenstates to nuclear displacement\cite{Hatomura2021,Pozsgay2024,Karve2025}—has recently emerged as a sensitive geometric tool for detecting chaos. However, its application to chemical reactivity, where vibrational modes serve as natural perturbative coordinates, remains nascent.

In this work, we introduce a chaos-diagnostic framework that integrates multireference quantum chemistry, AGP diagnostics, and RMT to investigate proton tunneling. Using the fundamental formation of H$_3^+$ and the H$_5^+$ cluster as representative model systems, we reveal that the Transition State (TS) acts as a dynamical bottleneck where chaos is notably suppressed. This ``integrable protection'' at the TS preserves quantum coherence, with direct implications for refining kinetic models of interstellar and planetary ionospheres\cite{Moore2019,Herbst2021}.

 Stationary points for the $H_{2} + H^{+} \rightarrow H_{3}^{+}$ and $H_{3}^{+} + H_{2} \rightarrow H_{5}^{+}$ reactions were optimized at the DFT/B3LYP/aug-cc-pVTZ level using Gaussian 09\cite{Gaussian09,augccpvtz}. Transition states were confirmed via QST3 algorithms and Intrinsic Reaction Coordinate (IRC) calculations\cite{QST3,IRC}. While high-accuracy benchmark potential energy surfaces exist for these fundamental ions \cite{Mizus2019,Xie2005}, the B3LYP topology provides a computationally efficient baseline sufficient for establishing the geometric chaos diagnostics presented here.To construct the Adiabatic Gauge Potential (AGP), we computed electronic eigenstates using state-averaged CASSCF/NEVPT2 within the PySCF framework\cite{PySCF}. We employed minimal active spaces of (2e,2o) for $H_{3}^{+}$ and (4e,4o) for the $H_{5}^{+}$ cluster. We explicitly acknowledge that while these active spaces are smaller than full-valence benchmarks, they capture the essential multi-reference character of the delocalized proton bridge required to test the AGP framework while maintaining tractability across extensive vibrational displacement scans ($\lambda_k \in [-0.5, +0.5]$). The AGP matrix elements, $A_{mn}^{(\lambda)} = \langle \Psi_m | \partial_{\lambda} H | \Psi_n \rangle / (E_n - E_m)$, were computed numerically along mass-weighted normal modes. To validate the chaos detection, we analyzed the nearest-neighbor level spacing statistics ($P(s)$) and the mean level spacing ratio ($\langle r \rangle$) of the vibronic Hamiltonian. Note that these statistics were computed on the generalized vibronic manifold without symmetry unfolding; thus, our reported $\langle r \rangle$ values serve as a macro-statistical diagnostic of global state mixing rather than a rigorous derivation of symmetry-resolved Wigner-Dyson statistics. Semiclassical tunneling probabilities were calculated using a mode-projected 1D WKB approximation along the CASSCF potential energy profiles. This reduction reduces the full-dimensional scattering problem to a transmission coefficient $T(\lambda_k)$, allowing us to isolate the specific contribution of individual vibrational modes to the "gating" mechanism..


We analyzed the full vibrational spectra of H$_3^+$ and H$_5^+$ at key reaction points—reactant, transition state (TS), and product—using normal mode displacement vectors (Figures S.1–S.4). This anchors our quantum chaos study in physically meaningful nuclear motions. H$_3^+$ exhibits three principal modes ranging from low-frequency collective motions to high-frequency H–H stretches, while H$_5^+$ comprises nine distinct modes including localized stretches, collective bends, and delocalized proton-bridging vibrations at the TS. High-frequency stretches primarily drive proton transfer, serving as key nuclear coordinates in hydrogen-bonded systems,\cite{Sitnitsky2023,Zhang2023,Tennyson2019} whereas lower-frequency modes modulate vibrational energy redistribution and chaotic dynamics. Mapping AGP norms and tunneling rates along these modes reveals vibrational channels that selectively enhance or suppress coherent tunneling.


To investigate the interplay between vibrational dynamics, quantum chaos, and tunneling, we performed systematic mode-resolved analysis using adiabatic gauge potential (AGP) norms and spectral statistics. The AGP norm $|A(\lambda)|$ provides a quantitative diagnostic for local quantum chaos: large values reflect strong electronic state mixing and loss of integrability, while small values signal modes that preserve quantum coherence.

\begin{figure}[htb]
\centering
\includegraphics[width=0.9\textwidth]{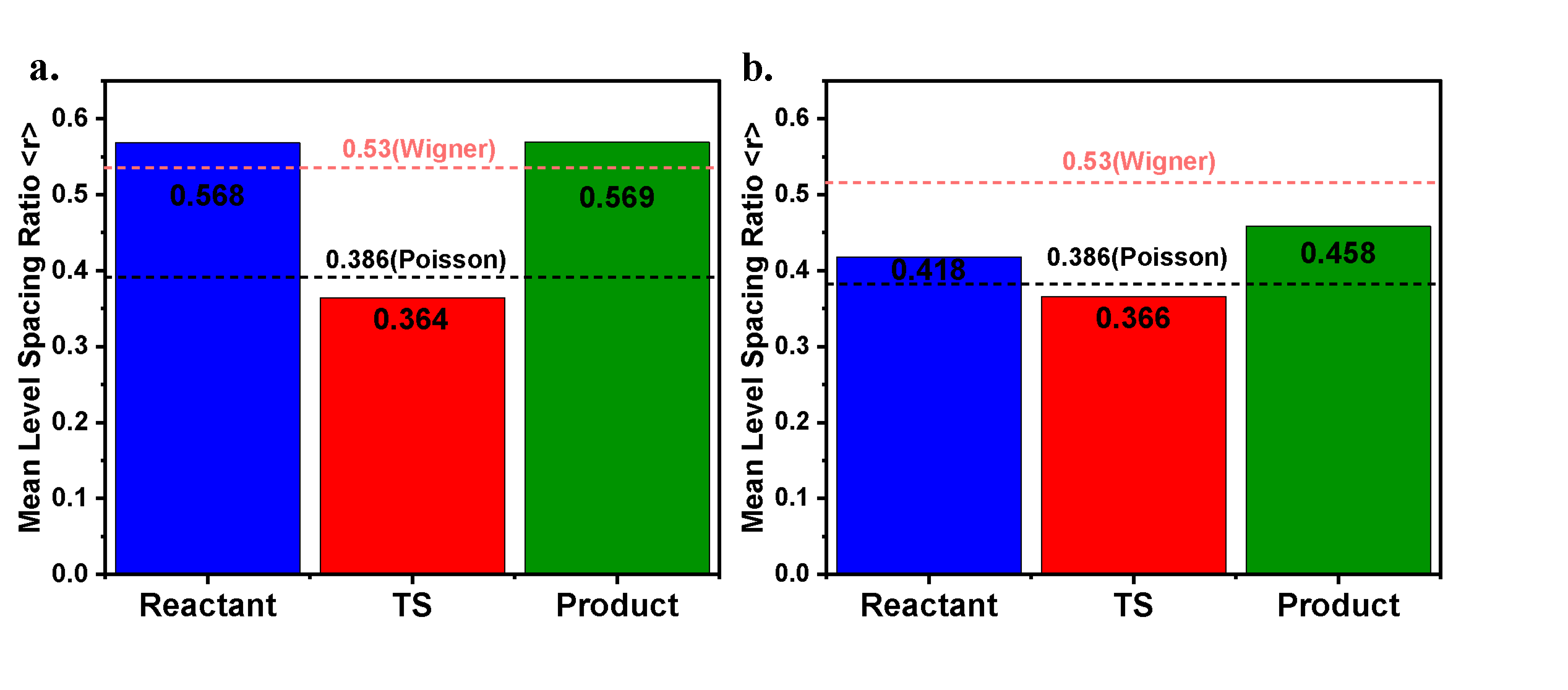}
\caption{(a) Mean level spacing ratio ($\langle r\rangle$) for  H$_2$ + H$^+$ $\rightarrow$ H$_3^+$ , (b) Mean level spacing ratio ($\langle r\rangle$) for H$_3^+$ + H$_2$ $\rightarrow$ H$_5^+$, }
\label{fig:H3_level_spacing_and_AGP}
\end{figure}

As illustrated in Figure~\ref{fig:H3_level_spacing_and_AGP}a, both reactant and product reaches GOE-like spectral statistics ($\langle r \rangle \approx 0.53$), indicative of quantum chaotic behavior. But interestingly the TS exhibits lower level spacing ratio indicative of integrable behavior($\langle r \rangle \approx 0.364$). This observation is paralleled in the literature on Anderson and many-body localization, where spectral statistics tend toward sub-Poisson values when eigenstates become exponentially localized and system dynamics are effectively one-dimensional~\cite{Alt2021,Falcao2022,Herrera2017,DeRoeck2025}. At the TS, normal mode analysis confirms that nuclear motion is funneled along a single (imaginary) mode, with all other degrees of freedom rendered effectively inert, leading to spectral sparsity and  suppression of level repulsion. Such behavior constitutes a dynamical signature of an absolute bottleneck for chaos and mixing, which is unique to the TS region.

\begin{figure}[htb]
\centering
\includegraphics[width=0.9\textwidth]{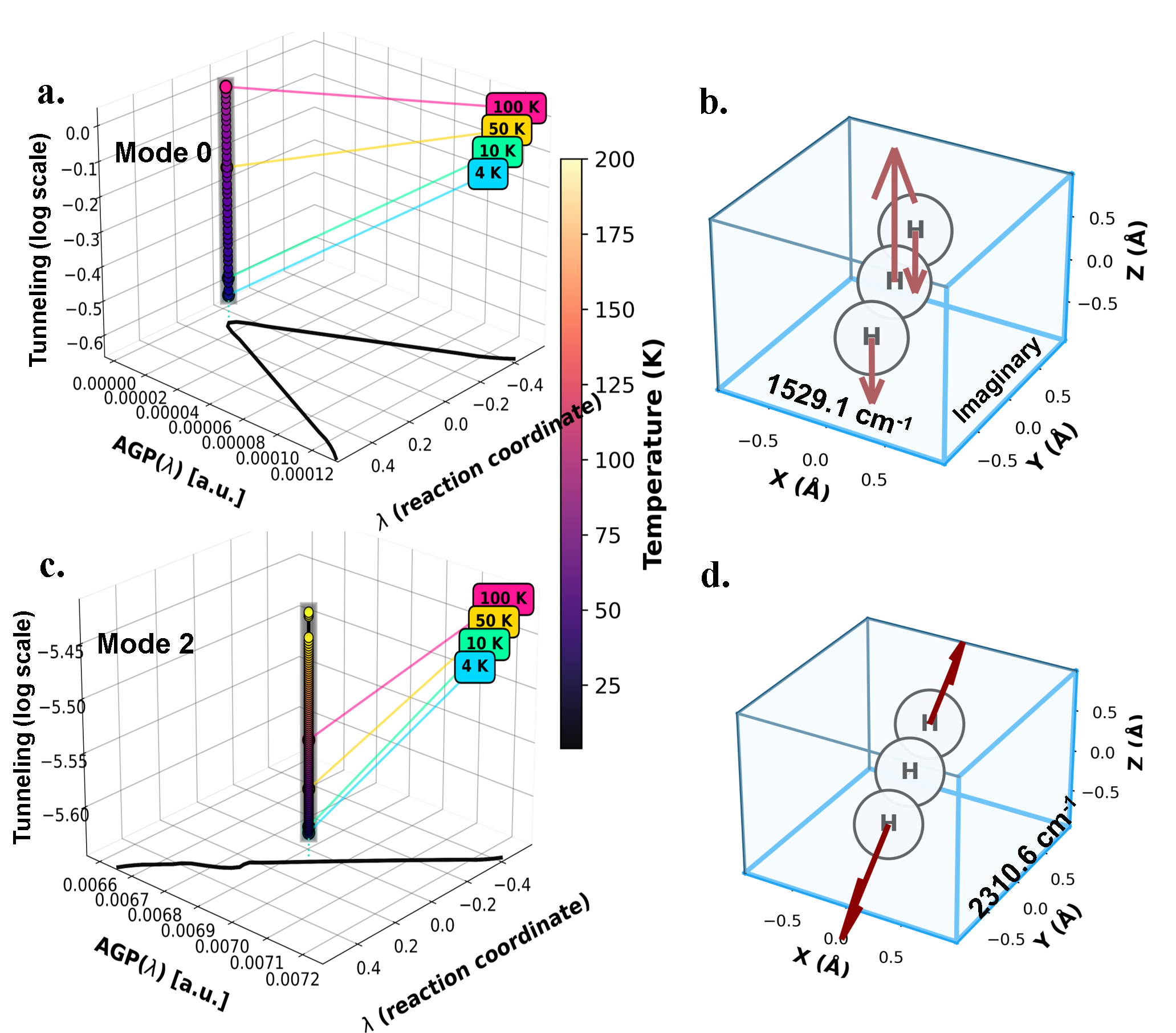}
\caption{{H\textsubscript{3}\textsuperscript{+} Mode-Resolved AGP and Tunneling}. 3D correlations of AGP, mode displacement $\lambda$, and tunneling probability (log scale) at the TS for selected vibrational modes. left: AGP, Tunneling correlation Per lambda, a. for mode 0, c for mode 2 Right: Vibrational mode structures and frequencies, b. Mode 0, d. Mode 2.}
\label{fig:H3_TS_modes}
\end{figure}

Figure~\ref{fig:H3_TS_modes} and Table~S.2 illustrate the profound mode-dependence of chaos-driven tunneling in the H$_2$ + H$^+$ $\rightarrow$ H$_3^+$ association reaction. For low-frequency, nearly integrable modes (e.g., Mode 0), $\log_{10}$AGP shows a flat, symmetric profile as a function of the displacement coordinate $\lambda$ (see Fig.~\ref{fig:H3_TS_modes}.a), and the corresponding tunneling probability remains high over a broad $\lambda$ range. Even with increasing temperature, these modes maintain robust quantum transmission, indicating that deviations from integrability minimal influence on their transport characteristics.

By contrast, higher-frequency modes (e.g., Mode 2) display steep, asymmetric log-AGP slopes (see Fig.~\ref{fig:H3_TS_modes}.c) and correspondingly sharp suppression of tunneling probability. For these modes, an order-of-magnitude increase in the AGP slope leads to a dramatic, three-orders-of-magnitude reduction in quantum transmission, which remains negligible even up to 200~K. This behavior establishes vibrationally-induced chaos as a direct, mode-selective barrier to quantum transport along the association coordinate.

The sign and magnitude of the AGP slope, $d[\log_{10}(\mathrm{AGP})]/d\lambda$, emerges as an effective fragility index: minor deviations from the symmetric TS geometry especially along high-frequency or asymmetric stretches activate chaos and rapidly suppress tunneling. Consequently, only vibrational motions aligned with the symmetry-adapted reaction coordinate promote efficient H$_3^+$ formation, whereas orthogonal distortions inhibit reaction through disorder-enhanced localization effects.

\begin{figure}[htb]
    \centering
    \includegraphics[width=0.9\textwidth]{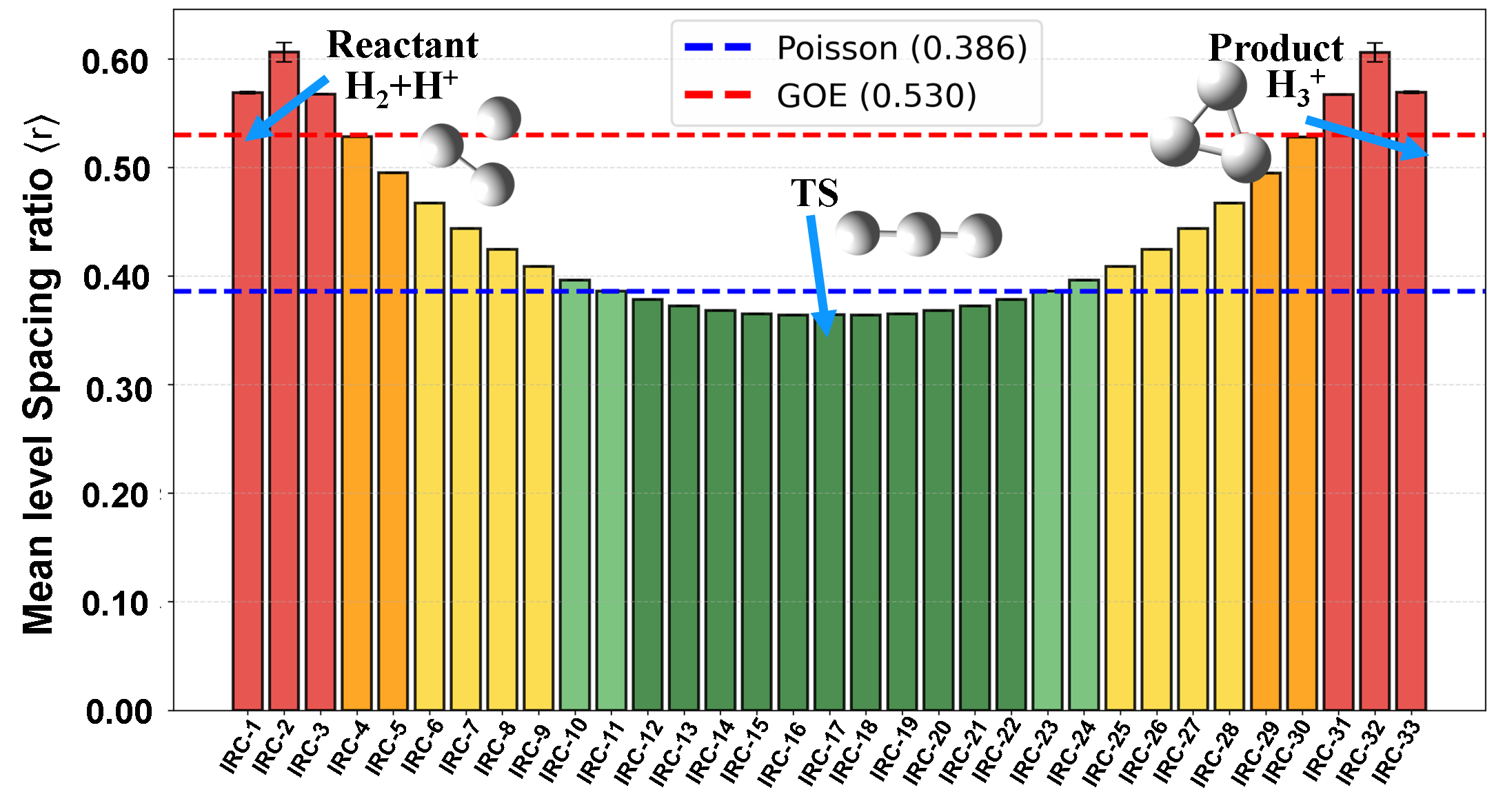}
    \caption{Distribution of $\langle r\rangle$ across all sampled geometries along the IRC for H$_2$ + H$^+$ $\rightarrow$ H$_3^+$ reaction from reactant to the product via the TS, showing the prevalence of chaotic (red columns) integrable (green columns) and weakly chaotic regimes (yellow columns).}
    \label{fig:H3_overall_chaos}
\end{figure}

We further analyzed the evolution of spectral chaos  along the intrinsic reaction coordinate (IRC) for the H$_2$ + H$^+$ $\rightarrow$ H$_3^+$ reaction, sampling 33 geometries from reactants to product through the TS. The reactant and product regions exhibit chaotic behavior, with average level spacing ratios, $\langle r \rangle$, above the GOE value (\(\sim 0.53\)), while the TS lies within an integrable or near integrable regime ($\langle r \rangle \leq 0.38$), indicating localized vibrational dynamics and suppressed level repulsion (Figure \ref{fig:H3_overall_chaos}). A similar trend is observed for the H$_3^+$ + H$_2$ $\rightarrow$ H$_5^+$ IRC, where reactants and products show weakly chaotic behavior ($\langle r \rangle \approx 0.42$–$0.46$) and the TS again marks a significant drop in chaos approaching integrable limits. This conserved pattern highlights the universal role of the TS as a dynamical constraint for quantum order within ion–molecule association processes.

The detailed IRC analysis is provided in SI. We further resolved this structural dependence through mode-resolved vibrational analysis at key geometries(reactant, TS and the product). For each structure, we displaced along all normal modes $\lambda$, computed the AGP norm, and evaluated the corresponding tunneling rates. 

\begin{figure}[H]
    \centering
    \includegraphics[width=0.7\textwidth]{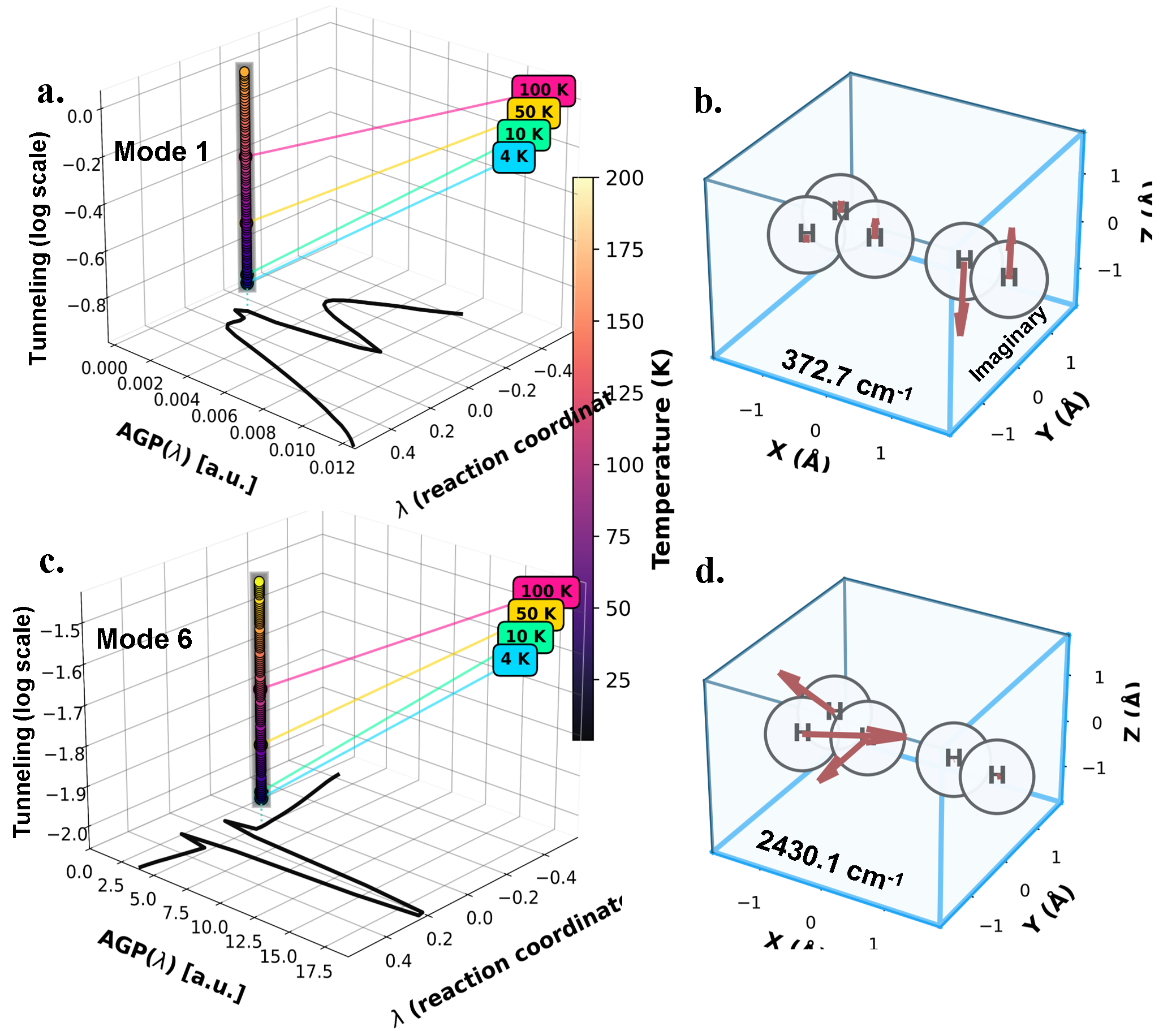}
    \caption{H\textsubscript{5}\textsuperscript{+} Mode-Resolved AGP and Tunneling. 3D AGP--$\lambda$--tunneling correlations (log scale) for representative {H\textsubscript{5}\textsuperscript{+}} modes at the transition state. Left: left: AGP, Tunneling correlation Per lambda, a. for mode 1, c for mode 6. Right: Vibrational mode structures and frequencies, b. Mode 1, d. Mode 6.}
    \label{fig:H5_modes1}
\end{figure}

Figure~\ref{fig:H5_modes1} and Supporting Figures~S.4–S.7 illustrate mode-resolved dynamics in H$_5^+$. Across both systems, low-frequency modes (0, 1) demonstrate small AGP variations and relatively high tunneling, consistent with integrable or weakly chaotic behavior and efficient quantum transmission, particularly at elevated temperatures. Intermediate-frequency modes show substantial diversity: Mode 3 exhibits nearly flat AGP yet vanishing tunneling, acting as a robust quantum bottleneck. Strong chaos activation emerges in Modes 5 and 6: Mode 5 shows a steep negative AGP slope that correlates with sharp tunneling suppression across all displacements, with only limited thermal recovery. Mode 6 displays highly asymmetric AGP with partial chaos-mediated leakage. High-frequency symmetric stretches (Mode 8) show elevated AGP and minimal transmission, with suppression persisting even under thermal excitation.
Temperature effects are most visible in low-chaos modes, where tunneling increases by over an order of magnitude at 200 K compared to 4 K. Conversely, Strong-chaos modes (5, 6, and 8) show limited thermal enhancement, reflecting robust chaos-induced barriers that persist under thermal excitation. These observations parallel those in H$_3^+$ and support a general principle wherein integrability facilitates efficient proton transport in these systems.
Mode-resolved  AGP profiles demonstrate that vibrationally induced chaos governs proton tunneling via eigenstate mixing. In \(\mathrm{H}_3^{+}\), low AGP slopes (\(\sim 2.6 \times 10^{-4}\)) enable robust tunneling (\(T_{\mathrm{base}} \approx 0.23\)), whereas high slopes (\(\sim 3.8 \times 10^{-2}\), Mode 3) induce strong suppression in mid-frequency modes—Mode 2 exhibits tunneling probabilities reduced to \(2.3 \times 10^{-6}\), corresponding to a five-orders-of-magnitude suppression compared to low-AGP modes. Remarkably, Mode 3 (highest frequency, steepest AGP) shows enhanced tunneling (\(T \approx 0.33\) at 4 K, rising to 0.72 at 200 K), revealing a non-monotonic AGP-tunneling relationship. In \(\mathrm{H}_5^{+}\), near-flat AGP modes show up to 8-fold thermal enhancement saturating near unity at 200~K, while chaotic modes remain suppressed, confirming that integrability—rather than barrier height—controls reactivity. The TS itself displays dual protection through Poisson-like spectral statistics and flat AGP landscapes, contrasting with chaotic reactants/products (\(\langle r \rangle \approx 0.54\)) suppress tunneling through mode localization. Detailed mode-by-mode analysis is provided in the supporting information. 

This AGP-chaos-tunneling framework provides a scalable diagnostic tool applicable to diverse hydrogen-bonded systems. Our analysis reveals a stark dynamical transition from chaotic reactant basins ($\langle r \rangle \approx 0.53$) to an integrable transition state ($\langle r \rangle \approx 0.36$), creating a ``protected'' funnel for proton transfer. By defining a ``fragility index,'' we demonstrate that steep AGP slopes correlate with a $10^5$-fold suppression in tunneling probabilities, offering a precise, data-driven metric for refining atmospheric and astrochemical models. Future work will extend these insights to asymmetric proton transfer and coupled vibrational dynamics, leveraging additional chaos descriptors to comprehensively capture chaos-induced tunneling control.

\section*{Author contributions}
SD, SP and KP conceptualized the work, performed the calculations, analyzed the data, and wrote the manuscript. JC and CS conceptualized the work, reviewed and edited the manuscript.

\section*{Conflicts of interest}
There are no conflicts to declare.

\section*{Acknowledgement}
The authors express their gratitude to IIST Thiruvananthapuram for the institutional support. We gratefully acknowledge the High Performance Computing Facility (HPS) at IIST for computational support. Language refinement was assisted by AI language models (OpenAI's GPT-4) for clarity and concision. 

\bibliography{references}   

\end{document}